\documentclass[
    ,final            
  ]
  {aipproc}

\usepackage{graphicx}
\usepackage{amsmath,amssymb}
\usepackage{color}

\layoutstyle{8x11single}

\begin{document}

\title{On the Origin of High-Energy Cosmic Neutrinos}


\author{Kohta Murase}{
  address={Institute for Advanced Study, Princeton, New Jersey 08540, USA}
}

\begin{abstract}
Recently, the IceCube collaboration made a big announcement of the first discovery of high-energy cosmic neutrinos.  Their origin is a new interesting mystery in astroparticle physics. The present multimessenger data may give us hints of connection to cosmic-ray and/or gamma-ray sources.  We look over possible scenarios for the cosmic neutrino signal, and emphasize the importance of multimessenger approaches in identifying the PeV neutrino sources and obtaining crucial clues to the cosmic-ray origin.  We also discuss some possibilities to study neutrino properties and probe new physics. 
\end{abstract}

\maketitle


\section{Introduction}
Neutrinos have served as an important messenger of the Universe.  The detection of MeV neutrinos from the supernova 1987A enabled us to confirm the basic picture of the death of massive stars.  Solar neutrinos have provided us with precious insights into fundamental properties of neutrinos.  High-energy neutrinos have also been of interest as special messengers for many years~\cite{Reines:1960we}.  First, their penetration power allows us to study high-energy physical processes such as dissipation in relativistic outflows, even if the outflows cannot be directly seen by electromagnetic observations.  Second, unlike gamma rays or charged particles, they can reach the Earth without significant attenuation or magnetic deflection in intergalactic space.  Identifications of neutrino sources enable us to study cosmic-ray (CR) accelerators and should help us solve a long-standing problem, the origin of CRs.  In particular, we may obtain crucial clues to the origin of ultrahigh-energy CRs (UHECRs) that are likely to come from extragalactic sources.   
       
Neutrinos interact with matter via weak interactions, which makes detecting high-energy neutrinos quite challenging.  However, after long-time efforts by experimentalists in different projects, promising events of cosmic neutrinos were finally discovered in the IceCube detector at the South Pole.  After the discovery of two shower-like events of PeV neutrinos~\cite{Aartsen:2013bka}, 26 additional events were found in the followup analysis~\cite{Aartsen:2013jdh}.  In 2014, the IceCube Collaboration reported further evidence for extraterrestrial neutrinos~\cite{Aartsen:2014gkd}, and the results may be summarized as follows.
\begin{enumerate}
\item
Based on the three year data, 37 events, including 9 track-like events, are found.  A spectral excess with respect to atmospheric backgrounds leads to significance $5.7\sigma$.  For a $E_\nu^{-2}$ spectrum, the isotropic diffuse neutrino intensity (for a single neutrino flavor) is $E_{\nu}^2\Phi_{\nu_i}\approx(0.95\pm0.3)\times{10}^{-8}~{\rm GeV}~{\rm cm}^{-2}~{\rm s}^{-1}~{\rm sr}^{-1}$, which is consistent with the previous analyses as well as independent results on muon neutrinos~\cite{Aartsen:2013eka}.
\item
Arrival directions of the observed neutrinos are consistent with isotropic distribution.  Any significant neutrino source has not been found yet.  
\item
For hard spectra like $E_\nu^{-2}$, a possible spectral break or cutoff is suggested at a few PeV energies.  Thanks to the Glashow resonance at 6.3~PeV for electron antineutrinos, a hard neutrino spectrum leads to overproduction of multi-PeV neutrinos~\cite{Laha:2013lka}.  However, it is not necessary for sufficiently large spectral indices, and the best-fit power law spectrum is $E_{\nu}^2\Phi_{\nu_i}\approx1.5\times{10}^{-8}{(E_\nu/0.1~{\rm PeV})}^{-0.3}~{\rm GeV}~{\rm cm}^{-2}~{\rm s}^{-1}~{\rm sr}^{-1}$.   
\end{enumerate}

The IceCube discovery has brought us new questions that are related to many open issues~\cite{Halzen:2013dva}.  What is the origin?  Are the main sources Galactic or extragalactic?  Is there any contribution from a single source?  Do neutrinos come from UHECR sources?  How are CRs accelerated?  Is the flavor ratio consistent with standard expectations?  Is their any contribution from nonastrophysical sources such as dark matter?  Is there any hint to new physics?  We demonstrate that multimessenger approaches using both gamma rays and neutrinos should be useful for addressing these issues.

\section{Cosmic-Ray Connection and Multimessenger Tests}
Whereas exotic possibilities such as dark matter decay have been discussed, we here focus on conventional astrophysical scenarios, where high-energy neutrinos are produced by CRs via hadronuclear (mainly $pp$) and/or photohadronic (mainly $p\gamma$) interactions.  The fact that the neutrino spectrum is at least extended to up 2~PeV sets a lower limit of the maximum CR energy per nucleon.  Since a neutrino typically carries $3-5$\% of the primary proton energy, we have 
\begin{equation}
E_\nu\sim0.04E_p\simeq2~{\rm PeV}~\varepsilon_{p,17}[2/(1+\bar{z})],
\end{equation}
where $\varepsilon_p={10}^{17}~{\rm eV}~\varepsilon_{p,17}$ is the proton energy in the cosmic rest frame and $\bar{z}$ is the typical source redshift.  Thus, neutrinos around a possible $\sim1-2$~PeV break are produced by protons with energies close to the {\it second/iron knee}~\cite{Murase:2008yt}.  Note that the CR energy per nucleon should exceed the {\it knee} energy at $3-4$~PeV.  Since the composition of Galactic CRs becomes heavier above the knee and the energy per nucleon is lower for nuclei with the same CR energy, naive adoption of the Galactic CR spectrum and composition leads to the extragalactic neutrino knee at $\sim100$~TeV.     

Arrival directions of IceCube neutrinos are consistent with isotropic distribution.  Then, the most natural possibility is an extragalactic origin, even though a fraction of the events may come from Galactic sources.  The mean diffuse neutrino intensity is evaluated by the line-of-sight integral, which can be approximated to be
\begin{eqnarray}
\sum_i(E_{\nu}^2\Phi_{\nu_i})\approx\frac{ct_Hf_z}{4\pi}\frac{3K}{4(1+K)}{\rm min}[1,f_{\rm mes}](E_pQ_{E_p})
\end{eqnarray}
where $t_H\simeq13.2$~Gyr, $f_{\rm mes}$ is the meson production efficiency, $Q_{E_p}$ is the differential CR energy budget at $z=0$, and $f_z$ is the redshift evolution factor~\cite{Waxman:1998yy,Murase:2012df}.  Note that $K=1$ and $K=2$ for $p\gamma$ and $pp$ interactions, respectively.  

\begin{figure}[tb]
\begin{minipage}{.45\linewidth}
\includegraphics[width=\textwidth]{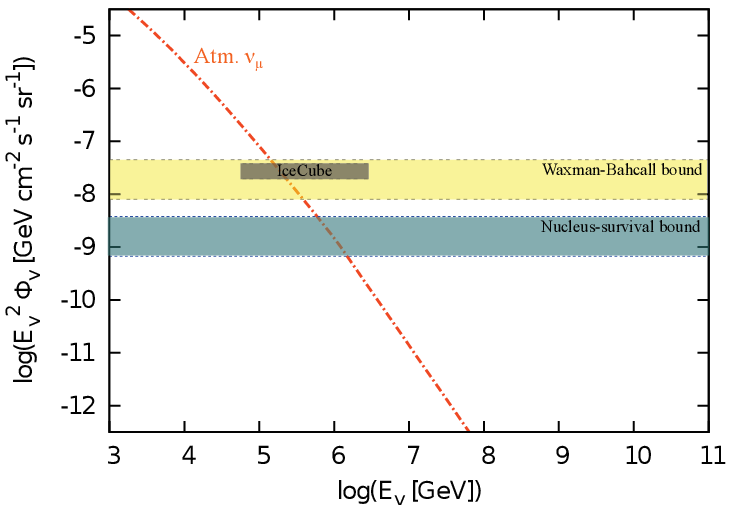}
\end{minipage}
\hfill
\begin{minipage}{.45\linewidth}
\includegraphics[width=\textwidth]{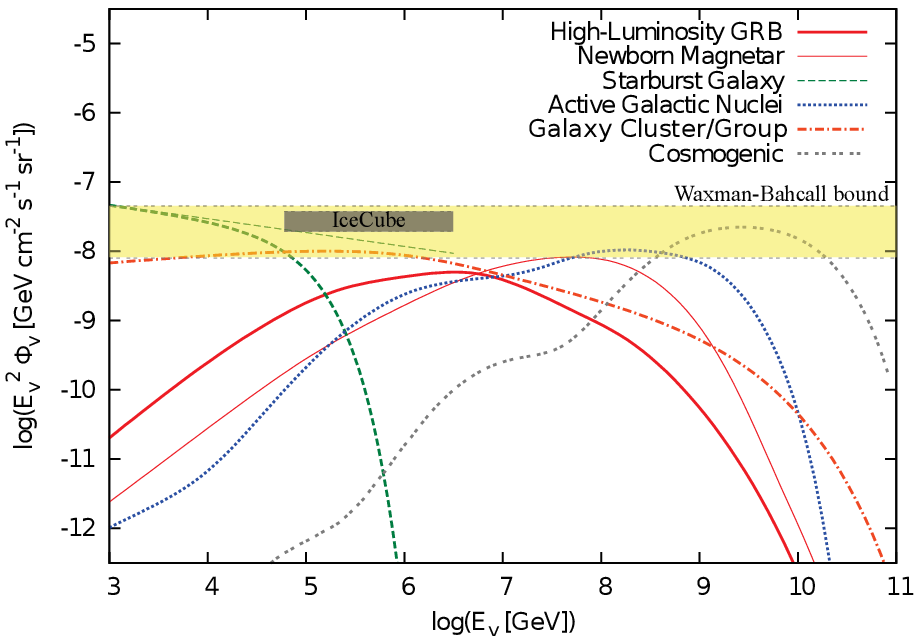}
\end{minipage} 
\caption{{\it Left panel}: The nucleon-survival and nucleus-survival landmarks for the sum of all three neutrino flavors.  Taken from Ref.~\cite{Murase:2010gj}.    
{\it Right panel}: Cumulative neutrino backgrounds predicted for various astrophysical sources.  Note that the models have large uncertainty in their predictions.  The numerical curves are taken from Refs.~\cite{Murase:2008sp,Murase:2009pg,Tamborra:2014xia,Murase:2014foa,Murase:2008yt,Takami:2007pp}. }
\end{figure}

In order to relate the observed diffuse CR and neutrino fluxes, it is useful to introduce a landmark flux so-called the Waxman-Bahcall bound~\cite{Waxman:1998yy,Bahcall:1999yr}.  This landmark is derived based on three assumptions: a) the meson production efficiency is taken to the formal limit of unity for semitransparent sources, (b) the injected CR spectrum is $E_pQ_{E_p}\propto$const. with the normalization based on UHECR data, and (c) intervening magnetic fields do not affect the observed flux of extragalactic CRs.  
The UHECR energy budget at ${10}^{19.5}$~eV is $E_pQ_{E_p}\approx0.6\times{10}^{44}~{\rm erg}~{\rm Mpc}^{-3}~{\rm yr}^{-1}$, leading to $\sum_i(E_{\nu}^2\Phi_{\nu_i})\simeq3\times{10}^{-8}f_z~{\rm GeV}~{\rm cm}^{-2}~{\rm s}^{-1}~{\rm sr}^{-1}$.  Although this does not exclude existence of hidden neutrino sources such that $f_{\rm mes}\gg1$, this nucleon-survival landmark is reasonable when the assumption (a) is justified.  Note that photodisintegration cross sections for nuclei are much higher than the photomeson cross section, so requiring nucleus survival leads to inefficient photohadronic neutrino production inside and/or outside the CR sources~\cite{Murase:2010gj}.
  
The measured diffuse neutrino flux matches the Waxman-Bahcall bound within uncertainty of redshift evolution models.  This fact indicates that the origin of the IceCube spectral excess may be related to UHECR sources or CR accelerators whose CR energy generation rate is not far from the UHECR one.  In the former case, for a $E_\nu^{-2}$ spectrum, $f_{\rm mes}\gtrsim1$ is required at 100~PeV proton energies.  But, to be consistent with the nondetection of $\gtrsim2$~PeV shower-like events, $f_{\rm mes}$ should be small above $\sim100$~PeV.  Alternatively, one may think of steeper spectral indices as in Figure~3 of Ref.~\cite{Murase:2010gj}.  Remarkably, the IceCube best-fit spectrum matches the nucleus-survival bound for $s_\nu=2.3$~\cite{Murase:2010gj}.  
In any case, the photomeson production efficiency cannot be too small if PeV neutrinos and UHECRs have the same origin~\cite{Yoshida:2014uka}.  If we consider transients, one source class could be the origin of CRs from GeV to ultrahigh energies~\cite{Katz:2013ooa,Waxman:2013zda}.  On the other hand, it is also possible to consider scenarios where PeV neutrino sources are not tightly related to UHECRs above ${10}^{19}$~eV.  As indicated in Equation~(1), PeV neutrinos are more closely related to CRs around the second knee, so it is important to reveal the origin of CRs around this energy.  

As we discuss below, the measured IceCube data can explained by some astrophysical scenarios, but we have neither identified any single source nor detected a sufficient number of neutrinos allowing statistical analyses.  To reveal their sources, multimessenger approaches combing gamma-ray and CR measurements are clearly important.  Whether neutrinos are produced via $pp$ or $p\gamma$ processes, gamma rays should be produced as well as neutrinos.  Interestingly, among gamma-ray sources that are also potential neutrino emitters, gamma-ray budgets of gamma-ray bursts (GRBs), radio-loud active galactic nuclei (AGN), and star-forming galaxies are $Q_\gamma\sim{10}^{44}~{\rm erg}~{\rm Mpc}^{-3}~{\rm yr}^{-1}$, $Q_\gamma\sim{10}^{44}-{10}^{46}~{\rm erg}~{\rm Mpc}^{-3}~{\rm yr}^{-1}$, and $Q_\gamma\sim{10}^{45}-{10}^{46}~{\rm erg}~{\rm Mpc}^{-3}~{\rm yr}^{-1}$, respectively, so their differential gamma-ray energy budgets are not far from the value required for the IceCube signal.  
Notably, the diffuse gamma-ray intensity has been measured by the {\it Fermi} satellite, and its origin has also been under debate.  Remarkably, the observed diffuse gamma-ray flux around 100~GeV is $E_{\gamma}^2\Phi_{\gamma}\sim{10}^{-7}~{\rm GeV}~{\rm cm}^{-2}~{\rm s}^{-1}$~\cite{Ackermann:2014usa}, which is comparable to the observed neutrino flux.  This fact indicates that the origin of the IceCube neutrino excess is related to gamma-ray sources that significantly contribute to the diffuse gamma-ray background.  The mean diffuse intensity of generated gamma rays is given by
\begin{equation}
E_{\gamma}^2\Phi_{\gamma}\approx\frac{ct_Hf_z}{4\pi}\frac{1}{1+K}{\rm min}[1,f_{\rm mes}](E_pQ_{E_p}),
\end{equation}
where \textit{model independently} we have
\begin{equation}
E_{\gamma}^2\Phi_{\gamma}\approx(4/K)(E_{\nu}^2\Phi_{\nu_i})|_{E_\nu=0.5E_\gamma}.
\end{equation}
For gamma rays, sufficiently high-energy gamma rays cannot avoid interactions with the cosmic microwave background and extragalactic background light, and they cause subsequent electromagnetic cascades~\citep[see][and references therein]{Murase:2012df}.  But these effects can be taken into account in the calculations.  First, if the neutrino and gamma-ray spectra are soft enough (as expected in $pp$ scenarios), only primary gamma rays are relevant.  Second, if the neutrino and gamma-ray spectra are hard (as expected in $p\gamma$ scenarios), cascades are essential but their spectrum is known to be nearly universal.  Regarding the \textit{Fermi} gamma-ray flux as an upper limit $E_{\gamma}^2\Phi_{\gamma}^{\rm up}$, we obtain $s_\nu\leq2+\ln[E_{\gamma}^2\Phi_\gamma^{\rm up}|_{100~{\rm GeV}}/(2E_{\nu}^2\Phi_{\nu_i}|_{E_{\nu}})]{[\ln(2 E_{\nu}/{100~{\rm GeV}})]}^{-1}$ for $K=2$.  Using $\sum_i(E_{\nu}^2\Phi_{\nu_i})=3\times{10}^{-8}~{\rm GeV}~{\rm cm}^{-2}~{\rm s}^{-1}~{\rm sr}^{-1}$ at $0.3$~PeV~\cite{Aartsen:2014yta}, the neutrino spectral index is constrained as~\cite{Murase:2013rfa}
\begin{equation}
s_\nu\lesssim2.185\left[1+0.265\log_{10}\left(\frac{(E_{\gamma}^2\Phi_\gamma^{\rm up})|_{100~{\rm GeV}}}{{10}^{-7}~{\rm GeV}~{\rm cm}^{-2}~{\rm s}^{-1}~{\rm sr}^{-1}}\right)\right].
\end{equation}
Thus, if PeV neutrino sources are GeV-TeV gamma-ray sources, their spectra must be hard enough.  Note that a pionic component of the diffuse Galactic emission has a much steeper index $s_\gamma\sim2.7$~\cite{FermiLAT:2012aa}.  We consider implications for astrophysical scenarios below. 

We note that the above multimessenger constraint is also applied to {\it unaccounted-for} Galactic sources, since the isotropic diffuse gamma-ray flux is a residual isotropic component obtained after subtracting known components including diffuse Galactic emission.  Noting that Galactic gamma-ray attenuation is negligible in the GeV-TeV range, the preliminary {\it Fermi} data already suggest $s_\nu\lesssim2.0$~\cite{Murase:2013rfa}, which also gives a stringent limit on some Galactic models.  
In addition, since sub-PeV gamma rays from Galactic sources can reach the Earth, constraints on the quasidiffuse gamma-ray intensity are quite powerful~\cite{Ahlers:2013xia,Supanitsky:2013ooa,Kalashev:2014vra}.  Ground-based detectors observing extended air showers, such as KASCADE and CASA-MIA, can detect not only hadronic showers caused by CRs but also electromagnetic showers caused by gamma rays. The large CR background can be discriminated by simultaneous detections of muons in muon-rich CR showers.  Advantages of surface array and water Cherenkov detectors are a large field of view and high duty cycle.  Interestingly, although many existing limits come from old measurements, the 0.1-10~PeV gamma-ray limits $E_{\gamma}^2\Phi_{\gamma}^{\rm up}\sim{10}^{-9}-{10}^{-8}~{\rm GeV}~{\rm cm}^{-2}~{\rm s}^{-1}~{\rm sr}^{-1}$ are comparable to the diffuse neutrino flux suggested by IceCube~\cite{Ahlers:2013xia}.

\section{Extragalactic Astrophysical Sources}
\subsection{Hadronuclear Production in Cosmic-Ray Reservoirs}
One of the main neutrino production channel is the inelastic $pp$ scattering.  This type of neutrino production is naturally expected for {\it cosmic-ray reservoirs}, where CRs escaping from their accelerators (e.g., supernovae and AGN) are confined in magnetized environments for a long time.  The $pp$ efficiency is estimated by
\begin{equation}
f_{pp}\approx n\kappa_{p}\sigma_{pp}^{\rm inel}ct_{\rm int},
\end{equation}
where $\kappa_{p}\approx0.5$ and $\sigma_{pp}^{\rm inel}\sim8\times{10}^{-26}~{\rm cm}^2$ at $\sim100$~PeV.  Also, $n$ is the target nucleon density, $t_{\rm int}\approx{\rm min}[t_{\rm inj},t_{\rm esc}]$ is the duration of interactions, $t_{\rm inj}$ is the CR injection time, and $t_{\rm esc}$ is the CR escape time.  One of the advantages is that $pp$ scenarios naturally predict a spectral break in the neutrino spectrum, which is indeed inferred by the IceCube data for spectra harder than $E_\nu^{-2.3}$.  Starburst galaxies and galaxy clusters/groups are most widely discussed, and a priori models can explain the cosmic neutrino signal within large astrophysical uncertainty.   

\begin{figure}
  \includegraphics[height=.3\textheight]{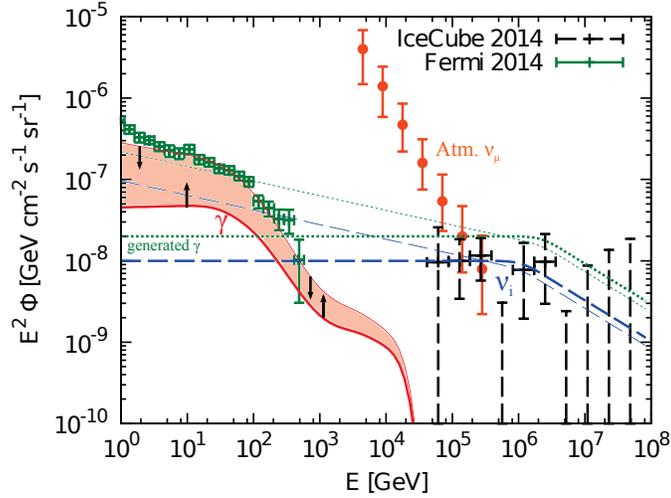}
  \caption{The allowed range in $pp$ scenarios that explain the cosmic neutrino signal observed by IceCube, which is indicated by the shaded area with arrows.  Taken from Ref.~\cite{Murase:2013rfa}.  The neutrino spectrum must be hard enough and the neutrino sources should give a significant contribution to the diffuse gamma-ray background.}
\end{figure}

\begin{description}
\item[Starburst Galaxies.---]
Supernova remnants have been believed to be the origin of Galactic CRs.  A special subset of star-forming galaxies, so-called starburst galaxies, are less numerous but more luminous, which are characterized by an enhanced star-forming activity in a very localized region.  No matter if such high star formation is triggered by galaxy mergers or by galactic bar instabilities, the birth of many massive stars leads to a lot of supernovae that inject CRs into these galaxies.  Since their magnetic fields are stronger than galaxies like the Milky Way, low-energy CRs are well confined and their escape is limited by advection.  Thanks to their high gas densities, $f_{pp}$ can be quite high and the observed neutrino flux level can be well explained as long as CRs can be accelerated above 100~PeV energies~\cite{Loeb:2006tw,Thompson:2006np,Murase:2013rfa,Tamborra:2014xia,Wang:2014jca}.  However, ordinary supernova remnants are not able to accelerate CRs up to this energy, so a lot of speculations have recently been made, including super-massive black hole activities~\cite{Tamborra:2014xia}, galaxy mergers~\cite{Kashiyama:2014rza}, super bubbles, interaction-powered supernovae~\cite{Murase:2010cu}, hypernovae~\cite{Sveshnikova:2003sa,He:2013cqa,Murase:2013rfa,Liu:2013wia}, transrelativistic supernovae and GRBs~\cite{Murase:2008mr,Budnik:2007yh,Wang:2007ya,Kashiyama:2012zn,Katz:2011zx}.  But it is highly unclear whether different populations lead to a smooth power-law spectrum and why the energy budget of rare transients is almost comparable to that of ordinary supernovae~\cite{Murase:2013rfa}.  At such high energies, diffusive escape of CRs would become relevant.  As a result, although properties of turbulence should be different from that in the Milky Way, a spectral break in the multi-PeV range is possible if the diffusion is close to the Bohm limit~\cite{Loeb:2006tw,Murase:2013rfa}.   
\item[Galaxy Clusters/Groups.---]
Galaxy clusters and groups, which contain many galaxies including AGN, are promising CR reservoirs~\cite{Berezinsky:1996wx}, and they are known to have magnetic fields with $\sim0.1-1~\mu{\rm G}$.  Relativistic jets of radio-loud AGN, weak jets of radio-quiet AGN, and disk-driven fast outflows from various types of AGN can be good CR accelerators~\cite{Murase:2013rfa,Tamborra:2014xia}, and inject CRs into the intracluster medium.  Sufficiently high-energy CRs may also be supplied by member galaxies and galaxy mergers.  In addition, one may expect contributions from CRs produced by accretion and cluster merger shocks that result from the structure formation of the Universe.  All CRs can be stored in clusters, and escaping CRs could contribute to the observed CR flux from the second knee to the ankle~\cite{Murase:2008yt,Kotera:2009ms}.  Given that groups and low-mass clusters can make significant contributions, the observed neutrino signal can be explained without contradicting existing gamma-ray limits~\cite{Murase:2008yt,Kotera:2009ms,Murase:2013rfa}.  A spectral break in the multi-PeV range, which can be caused by CR diffusion, has been expected~\cite{Murase:2008yt,Murase:2013rfa}.  However, gamma-ray emission of clusters has not been established yet, and the predictions are not certain enough.   
\end{description}

No matter what the CR reservoirs are, an important consequence is combing the neutrino and gamma-ray data leads to stringent {\it upper} limits on $s_\nu$ and {\it lower} limits on the contribution to the diffuse gamma-ray background~\cite{Murase:2013rfa}.  These results are largely independent of models, so useful multimessenger implications are obtained.  First, precise measurements of $s_\nu$ by sub-PeV neutrino observations with IceCube may support or exclude these $pp$ scenarios.  Large values of $s_\nu$ lead to overshooting of the diffuse gamma-ray background.  Second, understanding the sub-TeV diffuse gamma-ray background is important to reveal the origin of high-energy cosmic neutrinos.  If radio-loud AGN can account for more than $70-80$\% of the diffuse gamma-ray background, there is no room for such $pp$ neutrino sources that are responsible for the IceCube signal.  Third, TeV gamma-ray observations with Cherenkov detectors help us test $pp$ scenarios.  If gamma rays are detected, the sources should show hard spectra.  For bright and/or rare sources, stacking analyses can also be useful.

\subsection{Photohadronic Production in Cosmic-Ray Accelerators}
For extragalactic neutrino production, $p\gamma$ interactions are often more important.  A famous example is cosmogenic neutrino production in intergalactic space~\cite{Beresinsky:1969qj}, which is caused by CRs escaping from CR accelerators.  However, the typical energy of such off-source neutrinos is $\sim1-10$~EeV (unless a cutoff is introduced in the CR spectrum~\cite{Kalashev:2013vba}) and the spectrum is too hard to explain sub-PeV neutrino events.  Thus, this possibility is unlikely~\cite{Roulet:2012rv,Laha:2013lka,Aartsen:2013dsm}.  
  
For neutrino production inside CR accelerators, the most popular neutrino sources are GRBs and AGN, which are also known as powerful gamma-ray sources and popular candidate sources of UHECRs.  Observed nonthermal photons are usually attributed to synchrotron or inverse-Compton emission from relativistic electrons that are accelerated in relativistic jets.  Then, given that ions are accelerated as well, it is natural that sufficiently high-energy protons interact with ambient photons.  For a simple power-law target photon spectrum with $\propto\varepsilon_\gamma^{-\alpha}$, using the rectangular approximation, the energy-dependent $p\gamma$ efficiency is given by 
\begin{equation}
f_{p \gamma} \approx \frac{t_{\rm dyn}}{t_{p \gamma}} \simeq \frac{2 \kappa_\Delta \sigma_\Delta}{1+\alpha} \frac{\Delta \bar{\varepsilon}_{\Delta}}{\bar{\varepsilon}_{\Delta}} 
\frac{L_{\gamma}^b}{4 \pi r \Gamma_j^2 c \varepsilon_{\gamma}^b} {\left(\frac{\varepsilon_p}{\varepsilon_p^b} \right)}^{\alpha-1}, 
\end{equation}
where $t_{\rm dyn} \approx r/\Gamma_j c$ is the dynamical time, $t_{p \gamma}$ is the $p \gamma$ cooling time, $\sigma_\Delta \sim 5 \times {10}^{-28}~{\rm cm}^2$, $\kappa_\Delta \sim 0.2$, $\bar{\varepsilon}_{\Delta} \sim 0.3$~GeV, $\Delta \bar{\varepsilon}_{\Delta} \sim 0.2$~GeV, $L_\gamma^b$ is the luminosity at $\varepsilon_{\gamma}^b$, $r$ is the emission radius, and $\Gamma_j$ is the bulk Lorentz factor of jets.  The typical neutrino energy is given by
\begin{equation}
E_\nu^b\sim0.02 \Gamma_j^2 m_p c^2 \bar{\varepsilon}_\Delta{[\varepsilon_{\gamma}^b(1+\bar{z})]}^{-1}
\end{equation}
As a result, the neutrino spectrum often becomes harder than the injected CR spectrum.  Hence, in order to have soft neutrino spectra, one needs either of a soft CR spectrum or low-energy cutoff, or an appropriate shape of the luminosity function.  Another common feature of $p\gamma$ scenarios for relativistic sources is that the $p\gamma$ efficiency is sensitive to the Lorentz factor $\Gamma_j$, and we need good data of target photon spectra for solid theoretical predictions.    

\begin{description}
\item[Gamma-Ray Bursts and Supernovae.---]
GRBs are the most energetic explosive phenomena in the Universe, which potentially allows CRs to get accelerated up to ultrahigh energies.  Also, their gamma-ray energy generation rate is comparable to the UHECR energy budget, and they have been thought as one of the candidate sources of observed UHECRs.  GRBs have also been expected to be promising sources of PeV neutrinos~\cite{Waxman:1997ti}, and atmospheric backgrounds can be essentially removed by making use of time- and space-coincidence.  
In the case of prompt emission of classical high-luminosity GRBs with $\Gamma_j={10}^{2.5}$, $\sim10-100$~PeV protons interacting with MeV photons lead to the typical neutrino energy $0.3$~PeV.  Using typical values, $L_\gamma^b = {10}^{51.5}~{\rm erg/s}$, $\varepsilon_{\gamma}^b = 1$~MeV, and $r={10}^{14.5}$~cm, the typical $p\gamma$ efficiency is estimated to be $f_{p \gamma} \simeq 0.01$ although multipion production enhances $f_{p \gamma}$ by a factor of $\sim 3$.  
Due to strong meson cooling, having a multi-PeV break is possible~\cite{Cholis:2012kq,Winter:2013cla,Petropoulou:2014lja} which may be appealing in view of the spectral shape. but stacking searches by IceCube already set the limits of $E_{\gamma}^2\Phi_{\nu,i}\lesssim{10}^{-9}~{\rm GeV}~{\rm cm}^{-2}~{\rm s}^{-1}~{\rm sr}^{-1}$~\cite{Abbasi:2012zw}, so the simple standard scenario for high-luminosity GRBs cannot explain the IceCube signal~\cite{Laha:2013lka,Liu:2012pf}.  UHECR production may also be possible during GRB afterglows, where EeV neutrino production is expected~\cite{Waxman:1999ai,Dai:2000dj,Dermer:2000yd}. However, as in the argument for cosmogenic neutrinos, it is unlikely that afterglow neutrinos account for the observed IceCube signal~\cite{Murase:2007yt,Razzaque:2013dsa}.  Note that these conclusions do not mean that GRBs are not the main sources of UHECRs~\cite{He:2012tq,Li:2011ah,Hummer:2011ms}.  To cover all the relevant parameter space, more data are required.   

However, this is not the end of the story.  In recent years, various classes of interesting transients have been discovered.  Transrelativistic supernovae that are associated with low-luminosity GRBs show intermediate features of GRBs and supernovae.  There are classes of luminous or energetic supernovae, such as super-luminous supernovae and hypernovae.  For example, low-luminosity GRBs are much more common than high-luminosity GRBs~\cite{Liang:2006ci}, and their CR energy budget may be comparably important.  Low Lorentz factors allow efficient neutrino production, and a calculation apparently matches the IceCube spectral excess~\cite{Murase:2006mm,Gupta:2006jm,Murase:2013ffa} without contradicting single source limits~\cite{Ahlers:2014ioa}.  In addition, although radiation constraints forbid efficient CR acceleration in powerful hidden jets inside a star, low-power GRBs are favorable as sources of orphan and precursor neutrinos~\cite{Murase:2013ffa}.  Ultralong GRBs, if they come from blue super-giants (but see discussion in Ref.~\cite{Zhang:2013coa}), could contribute to the diffuse neutrino flux with rather steep spectral indices~\cite{Aartsen:2014yta}.  A multi-PeV spectral break can come from the maximum CR energy, meson cooling and neutrino attenuation in the stellar material~\cite{Murase:2013ffa}.  However, we note that the models are highly uncertain simply because such transients are not well understood.  The message here is that we should keep up good GRB monitors especially in the hard X-ray range.  More GRB samples are needed to perform stacking analyses on such transients, and understanding them gives us clues to connections between GRB and supernovae.  

\begin{figure}[t]
\begin{minipage}{.45\linewidth}
\includegraphics[width=\textwidth]{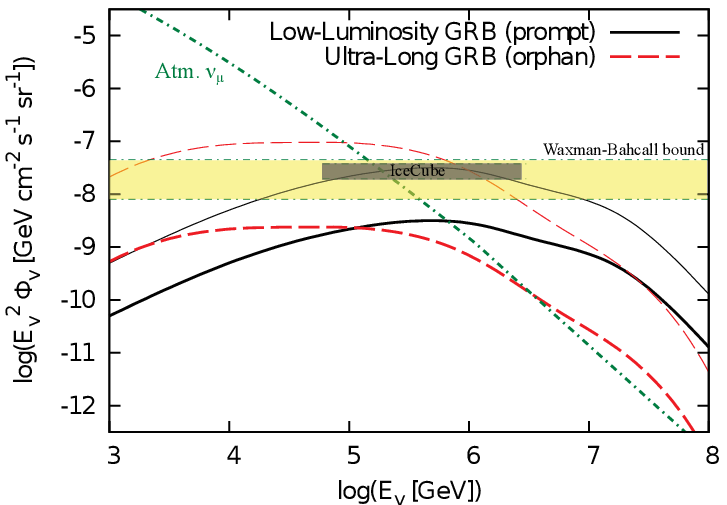}
\end{minipage}
\hfill
\begin{minipage}{.45\linewidth}
\includegraphics[width=\textwidth]{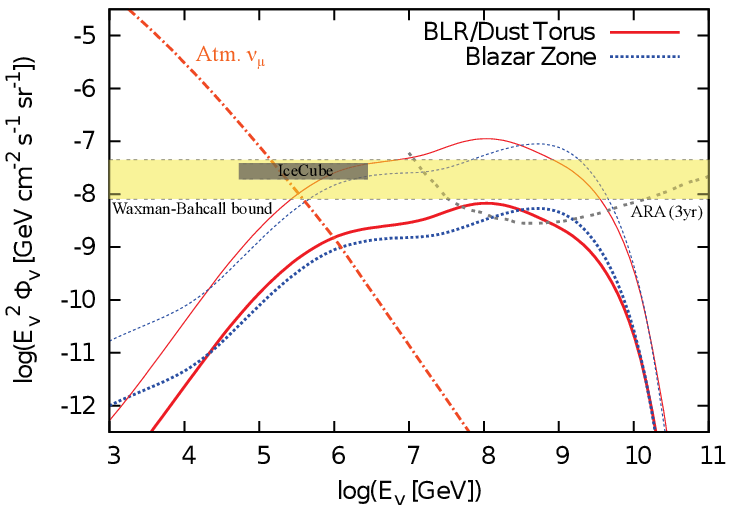}
\end{minipage} 
\caption{{\it Left panel}: The cumulative neutrino backgrounds from low-luminosity and ultralong GRBs.  For low-luminosity GRBs, we consider $(\xi_{\rm acc}/10)(\rho/500~{\rm Gpc}^{-3}~{\rm yr}^{-1})=0.2-2$.  The spectrum of low-luminosity GRBs was originally taken from Ref.~\cite{Murase:2006mm}.  For ultralong GRBs, we consider $(f_{\rm cho}/10)(\rho/1~{\rm Gpc}^{-3}~{\rm yr}^{-1})=0.1-4$, where $f_{\rm cho}=40$ is based on the requirement that the true rate is less than the hypernova rate.  Taken from Ref.~\cite{Murase:2013ffa}, but Figure~3 is modified to demonstrate astrophysical uncertainty.  
{\it Right panel}: The cumulative neutrino backgrounds from radio-loud AGN~\cite{Murase:2014foa}, but the latest IceCube data are indicated.}
\end{figure}

\item[Active Galactic Nuclei.---]
AGN host super-massive black holes, and they have been the most popular candidate sources of UHECRs.  Various scenarios have been discussed, including CR production in the vicinity of the black hole or accretion disk, inside jets, at hot spots and cocoon shocks, and in lobes and bubbles.  In particular, powerful jets of radio-loud AGN including blazars have been widely discussed as potential sites of high-energy neutrino production~\cite{Mannheim:1995mm,Atoyan:2001ey,Atoyan:2002gu}.  For AGN jets with $\Gamma_j=10$ and a synchrotron peak $\varepsilon_{\gamma}^b =10$~eV, the typical neutrino energy is expected to be $E_\nu\sim30$~PeV.  For $L_\gamma^b={10}^{45.5}~{\rm erg/s}~$ and $r={10}^{17.5}$~cm, one has $f_{p \gamma}\sim0.9\times{10}^{-3}{(E_p/1.6\times{10}^{18}~{\rm eV})}^{1/2}$.  On-axis radio-loud AGN, so-called blazars, are composed of high-luminosity (quasar-hosted blazars) and low-luminosity (BL Lac objects) classes.  In the former class, external photon fields provided by the broadline region and dust torus become more important than synchrotron photons as targets for photomeson production~\cite{Murase:2014foa,Dermer:2014vaa}.  The diffuse neutrino flux is dominated by such luminous blazars, and a power-law CR spectrum leads to a hard neutrino spectrum with a low-energy cutoff feature around PeV energies, which is inconsistent with the existing neutrino data.  Thus, although more complicated models are possible~\cite{Dermer:2014vaa,Tavecchio:2014iza}, the simple standard scenario for radio-loud AGN jets has difficulty in explaining the IceCube signal.  It is worthwhile to mention that some optimistic pre-IceCube predictions are already ruled out~\cite{Waxman:2013zda}.  
However, this does not mean that radio-loud AGN are disfavored as the origin of UHECRs.  Interestingly, future detectors such as the Askaryan Radio Array can detect EeV neutrinos even in such conservative cases, allowing us to indirectly test the possibility of UHECR acceleration.  One of the predictions is that the arrival directions of neutrinos correlate with luminous blazars detected by {\it Fermi}~\cite{Murase:2014foa}.
Note that $pp$ interactions are unlikely to be relevant in the blazar zone~\cite{Atoyan:2002gu}. They are also insignificant in large scale jets.  They could be relevant in the vicinity of the black hole~\cite{Anchordoqui:2004eu,AlvarezMuniz:2004uz,Tjus:2014dna} but $p\gamma$ interactions can compete~\cite{Stecker:1991vm}, depending on properties of the accretion disk.  Such AGN core models can be viable~\cite{Stecker:2013fxa,Winter:2013cla} but uncertainty is much larger.  Gamma rays may not escape either~\cite{Dermer:2014vaa}, so more studies are needed to test these possibilities.  
\end{description}

Multimessenger tests based on gamma-ray measurements are applicable but more complicated.  First, the $p\gamma$ process is efficient only for sufficiently high-energy CRs, so GeV-TeV gamma rays are produced via cascades and low-energy pionic gamma rays do not directly contribute to the diffuse gamma-ray background.  More seriously, the tight connection is not expected in $p\gamma$ sources like GRBs and AGN, since target photons for $p\gamma$ reactions often prevent GeV-PeV gamma rays from leaving the sources.

\section{Galactic Astrophysical Sources}
One of the basic questions is if the sources are Galactic or extragalactic.  
Although the extragalactic origin is the most natural, as demonstrated in Ref.~\cite{Ahlers:2013xia} (see also Ref.~\cite{Taylor:2014hya}), one could consider quasiisotropic emission produced by CRs confined in the Galactic Halo.  The circumgalactic density suggested from observations of dozens of nearby galaxies indicating a decreasing density profile~\cite{Werk:2014fza}, which is also consistent with theoretical expectations.  Then, for $s_\nu\sim2$, we typically expect $\sum_i(E_{\nu}^2\Phi_{\nu_i})\sim7\times{10}^{-9}~{\rm GeV}~{\rm cm}^{-2}~{\rm s}^{-1}~{\rm sr}^{-1}$, for a typical enhancement factor of past star formation~\cite{Ahlers:2013xia}.  
Interestingly, even if higher fluxes can somehow be achieved by taking more extreme values, this possibility can strongly be constrained by gamma rays.  First, {\it Fermi} gamma-ray constraints are applicable to isotropic emission, so $s_\nu\lesssim2.0$ will be needed~\cite{Murase:2013rfa}.  Second, there is a tension with sub-PeV gamma-ray limits obtained with two independent measurements, CASA-MIA and KASCADE.  Note that $\sim60$\% of the contributions come from regions within $\sim30$~kpc, which is enhanced by the realistic CR density gradient.  Thus, the present multimessenger data support the extragalactic origin of the IceCube signal, but more careful studies are needed to have a robust statement. 

\begin{figure}
\begin{minipage}{.45\linewidth}
\includegraphics[width=\textwidth]{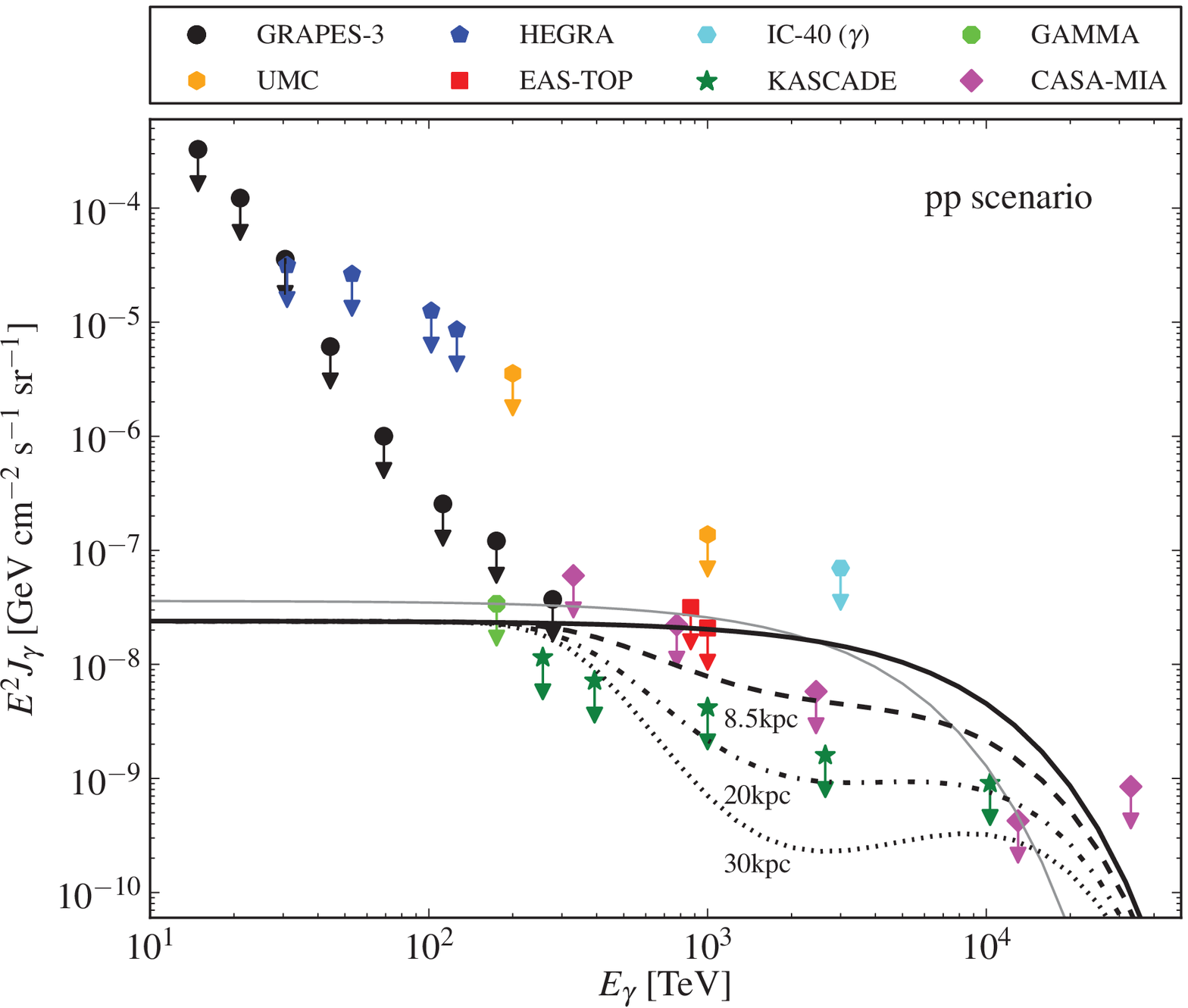}
\end{minipage}
\hfill
\begin{minipage}{.45\linewidth}
\includegraphics[width=\textwidth]{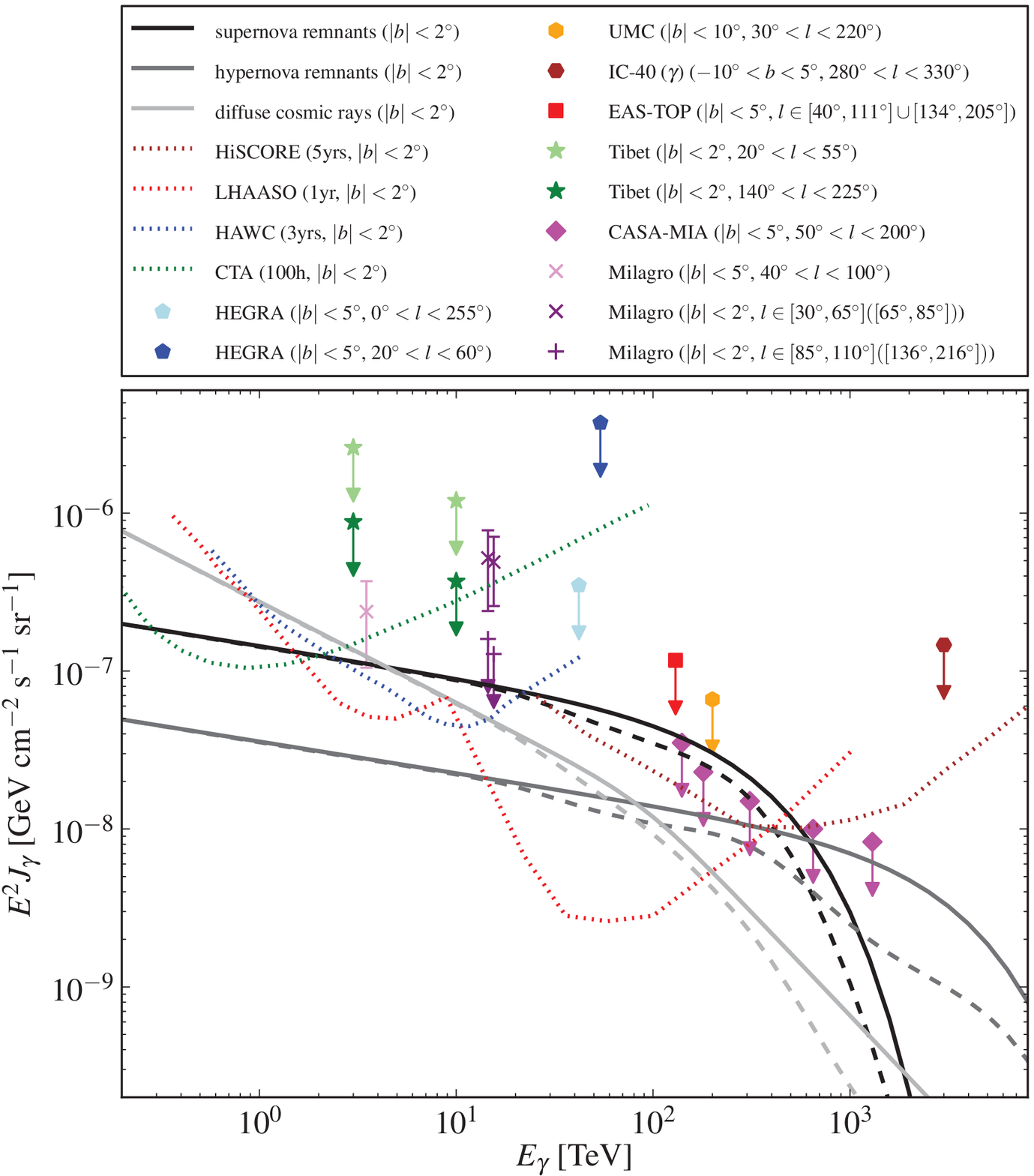}
\end{minipage} 
\caption{{\it Left panel}: The diffuse gamma-ray flux that corresponds to the diffuse neutrino flux in the quasiisotropic Galactic scenario.  An exponential cutoff at 6 PeV (for gamma rays) is assumed.  Gamma-ray attenuation is shown for 8.5 kpc, 20 kpc, and 30 kpc.  Taken from Ref.~\cite{Ahlers:2013xia} 
{\it Right panel}: Various measurements of the diffuse gamma-ray flux around the Galactic plane.  For comparison, theoretical curves for diffuse Galactic emission, unresolved supernova remnants, and unresolved hypernova remnants are shown. Taken from Ref.~\cite{Ahlers:2013xia}}
\end{figure}

\begin{figure}
  \includegraphics[height=.3\textheight]{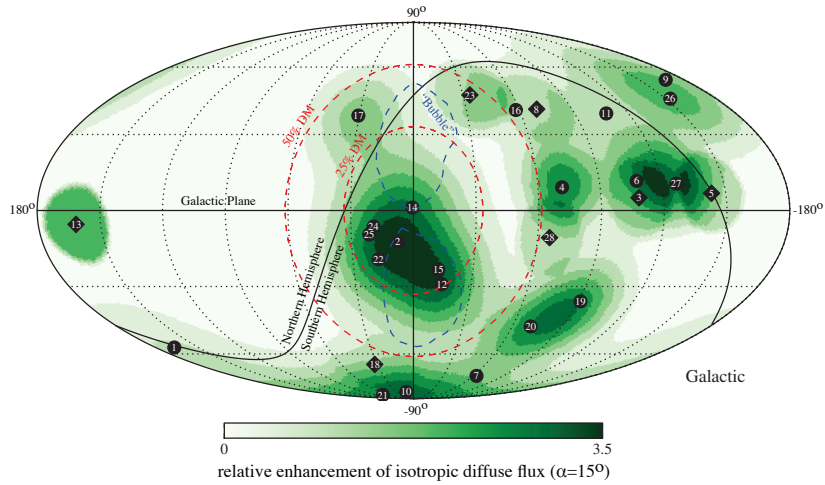}
  \caption{The fluctuation of the nonisotropic diffuse flux, where emission is extended with $15^\circ$ around each neutrino direction in the sky.  Taken from Ref.~\cite{Ahlers:2013xia}.}
\end{figure}

In addition, there are attempts that try to connect a fraction of the neutrino events to some Galactic sources.  Although none of them are statistically significant at present, some contributions from Galactic sources cannot yet be discarded and large scale anisotropy could exist~\cite{Anchordoqui:2014hia}.  
In principle, one may expect the Galactic to extragalactic transition in the neutrino spectrum, and the IceCube spectral excess may consist of a superposition of Galactic and extragalactic events especially below $\sim200$~TeV.  Galactic neutrino signals, if confirmed, may infer Galactic pevatorons, so it is interesting to seek event clustering.  Diffuse TeV-PeV gamma-ray limits are powerful, but gamma-ray surveys are biased in the Northern Hemisphere and they do not have significant overlaps with the observed neutrino events~\cite{Ahlers:2013xia}.   
\begin{description}
\item[Galactic Disk.---]
CRs make nonisotropic diffuse neutrino emission during their propagation in the interstellar medium, which can be regarded as guaranteed emission~\cite{Stecker:1978ah,Berezinsky:1992wr}.  However, the expected spectrum with $s_\nu\sim2.7$ is too soft, and it has been expected to have the neutrino knee around 100~TeV~\cite{Kachelriess:2014oma,Joshi:2013aua}.  Moreover, there is no significant event clustering along the Galactic plane within latitude $2.5^\circ$ although an extension to larger angular scales could include up to 13 events within uncertainties~\cite{Ahlers:2013xia}.  The fluctuation is difficult to be attributed to Galactic diffuse emission from propagating CRs~\cite{Ahlers:2013xia}.  
\item[Supernova/Hypernova Remnants and Pulsar Wind Nebulae.---]
Supernova remnants and pulsar wind nebulae are distributed in the Galactic disk and their emissions are clustered within latitude $\sim2^\circ$.  But signifiant event clustering around the Galactic plane is not seen at present~\cite{Ahlers:2013xia}.  However, nearby sources in the local spiral arm of the Milky Way may appear at higher latitudes~\cite{Neronov:2013lza}.  There is a fluctuation at longitudes $\sim240^\circ$, although no apparent neutrino clustering is found in the opposite direction~\cite{Ahlers:2013xia}.  Very-high-energy gamma-ray observations should give us useful tests.  Although supernova remnants are believed to be responsible for CR protons below the knee, deeper diffuse gamma-ray limits will imply that not all young supernova remnants do accelerate protons up to the CR knee, which is consistent with TeV gamma-ray observations of nearby remnants.  A possible contribution from hypernova remnants and unidentified TeV sources~\cite{Fox:2013oza} can also be tested by sub-PeV gamma-ray measurements~\cite{Ahlers:2013xia}.   
\item[Galatic Center and Fermi Bubbles.---]
The Galactic center is a complicated region where many gamma-ray sources exist, and there is a fluctuation around this region~\cite{Ahlers:2013xia,Razzaque:2013uoa}.  The super-massive black hole in the Galactic center has also been suggested as a potential CR accelerator and neutrino emitter~\cite{Bai:2014kba}.  Fermi bubbles, which may be driven by past super-massive black hole activities or starburst winds, are observed as an extended gamma-ray source with a hard spectrum.  It has been suggested that this emission is caused by $pp$ interactions of CRs that are injected from the Galactic center fort $\sim1-10$~Gyr~\cite{Crocker:2010dg}.  In this model, the neutrino flux suggested from fluctuations around the Fermi bubbles can be consistent with a a power index $s_\nu\approx2.2$~\cite{Ahlers:2013xia,Lunardini:2013gva}.  On the other hand, the leptonic emission model is also viable that seems more consistent with the recent indication of a high-energy gamma-ray cutoff~\cite{Fermi-LAT:2014sfa}.  Future gamma-ray observatories as HAWC should be able to test these possibilities~\cite{Ahlers:2013xia}.
\item[Gamma-Ray Binaries.---]
Gamma-ray binaries are composed of a compact object and a massive nondegenerate star, and some of high-mass X-ray binaries are classified as microquasars, where resolved and collimated radio features are seen.  The compact object can be a black hole, where jets can be powered by accretion from the companion star.  Alternatively, a collision between a relativistic pulsar wind and a stellar wind may lead to particle acceleration at the shock. Although no significant correlation with them is found at present, it has been argued that some of the observed neutrinos could come from one of the binaries, LS 5039~\cite{Anchordoqui:2014rca}. 
\end{description}

\section{Discussion and Summary}
The origin of high-energy cosmic neutrinos is a new mystery in astroparticle physics.  In this talk, we looked over various possibilities, focusing on astrophysical sources.  The observed neutrino signals are likely to include extragalactic contributions, even though some of them might come from Galactic sources.  Although heavy dark matter models are not excluded at present~\cite{Murase:2012xs,Feldstein:2013kka,Esmaili:2013gha,Bhattacharya:2014vwa,Zavala:2014dla,Rott:2014kfa}, connections to diffuse CR and gamma-ray data suggest that the origin of IceCube neutrinos is likely to be related to extragalactic CRs and/or gamma-ray sources contributing to the diffuse gamma-ray background.  It is also supported by the fact that some pre-IceCube calculations for extragalactic astrophysical sources can explain the IceCube spectral excess~\cite{Loeb:2006tw,Murase:2008yt,Murase:2006mm}.  Among various interpretations, CR reservoirs such as starburst galaxies and galaxy clusters seem appealing as the origin of PeV neutrinos.  In such $pp$ scenarios, the existence of a spectral break in the multi-PeV range was predicted and required to explain the IceCube data.  We obtained new powerful multimessenger constraints.  Measurements of $s_\nu$ by sub-PeV neutrino observations will give further tests for $pp$ scenarios, and large values of $s_\nu$~\cite{Aartsen:2013eka} may imply the existence of other components.  PeV neutrino sources should give a significant contribution to the diffuse gamma-ray background, so understanding the sub-TeV diffuse gamma-ray background is important.   
On the other hand, the situation is a bit complicated in $p\gamma$ scenarios including relativistic sources as UHECR accelerators.  The simple standard jet models for high-luminosity GRBs and radio-loud AGN are already disfavored as the origin of the IceCube spectral excess.  The remaining possibilities include low-power GRBs and AGN core models (including low-power AGN classes).  Although some a priori models may explain the data, they are quite uncertain.  Since gamma-ray attenuation and their cascades are often crucial, the multimessenger connection is less obvious, so that further theoretical and observational studies are needed. 
 
Galactic sources may give some contributions especially below $\sim200$~TeV, and can help the diffuse neutrino spectrum be steeper.  Possible subdominant neutrino sources like supernova/hypernova remnants are marginally consistent with present multi-TeV and sub-PeV gamma-ray limits.  On the other hand, quasiisotropic Galactic emission models are being constrained by the existing TeV and sub-PeV gamma-ray data.  We note that sky regions from which neutrino events are not much covered by air shower detectors.  This emphasizes the importance of having diffuse TeV-PeV gamma-ray monitors in the Southern Hemisphere.
 
One of the interesting question is if sources of PeV neutrinos are related to UHECR sources or not.  As in the Waxman-Bahcall bound, for a hard CR spectrum, the photomeson production efficiency has to be as large as $f_{\rm mes}\gtrsim1$, where a spectral break in the neutrino spectrum is required to avoid overproduction of multi-PeV neutrinos.  In this case, it is not easy for neutrinos produced inside UHECR accelerators (such as GRBs and AGN) to consistently explain PeV neutrinos and UHECRs simultaneously.  But, even so, transients are appealing as UHECR sources, and a flat CR spectrum in the wide energy range may be explained~\cite{Katz:2013ooa}.  It is also possible to have $f_{\rm mes}\sim0.01-0.1$ or even smaller values if environment magnetic fields are relatively strong, where steeper spectra are needed.  As in the nucleus-survival bound, this case would be more favorable if the UHECR composition is heavy.  Alternatively, one does not have to connect PeV neutrino sources to UHECR sources.  Although this case may not be as interesting as the former two, one can still expect interesting connection to gamma-ray sources.  

In this talk, we discussed implications based on the observed neutrino spectrum and arrival directions.  Flavor studies, through analyzing track-like, shower-like events and possible $\nu_\tau$ signatures, should also give us insightful information on the origin.  Theoretically, if neutrinos mainly come from pion decay, $\nu_e:\nu_\mu:\nu_\tau\approx1:1:1$ is expected in the cosmological baseline limit~\citep[e.g.,][]{Learned:1994wg,Beacom:2003nh}.  However, for muon-damped sources (that happen if muons strongly cool before they decay), one should have more track-like events as $\nu_e:\nu_\mu:\nu_\tau\approx0.57:1:1$~\cite{Kashti:2005qa,Blum:2007ie}.  If neutrinos mainly come from neutron decay, one has $\nu_e:\nu_\mu:\nu_\tau\approx2.5:1:1$.  Thus, flavor studies will enable to us learn source physics, although the ratio could be affected by astrophysical processes such as reacceleration of mesons and muons.  In addition, although $\nu_\tau$ signatures have not been found so far, observing the $\tau$ appearance should also be helpful.  

\begin{figure}
\begin{minipage}{.45\linewidth}
\includegraphics[width=\textwidth]{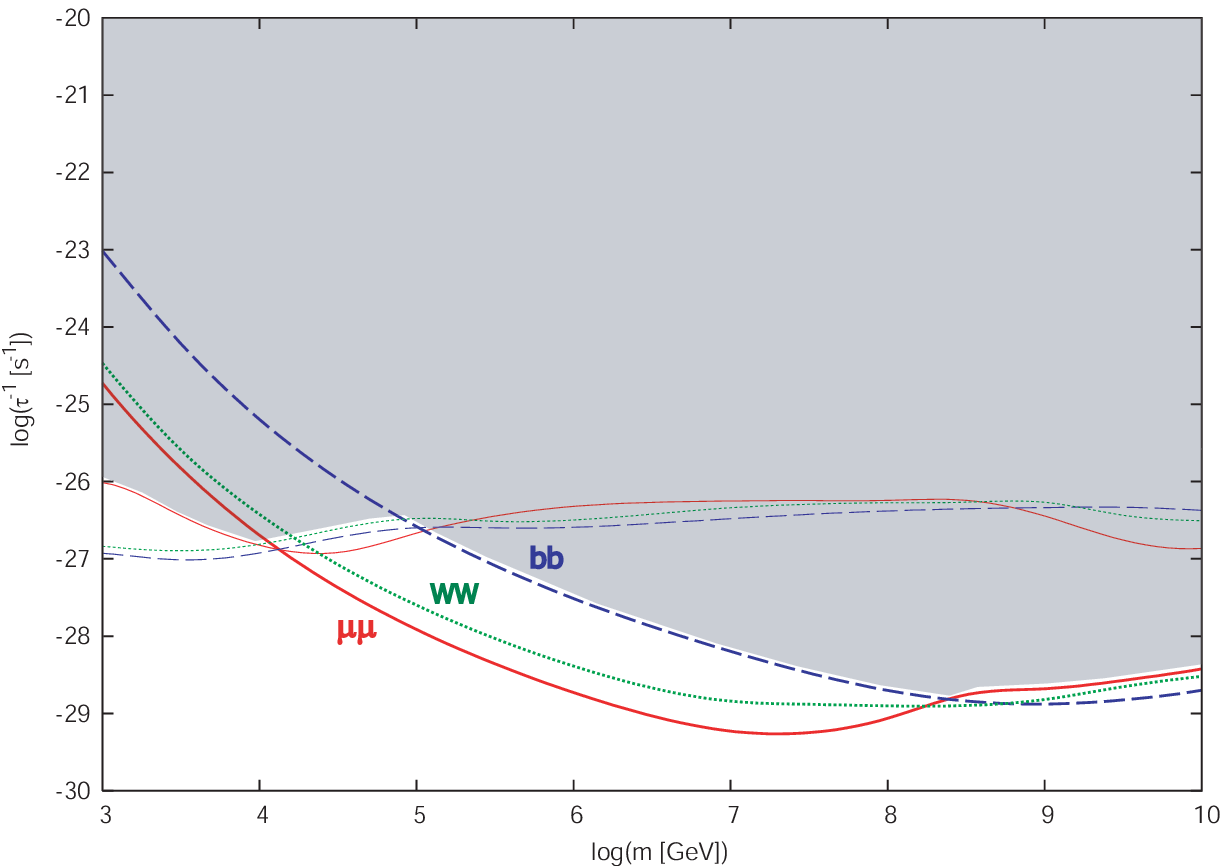}
\end{minipage}
\hfill
\begin{minipage}{.45\linewidth}
\includegraphics[width=\textwidth]{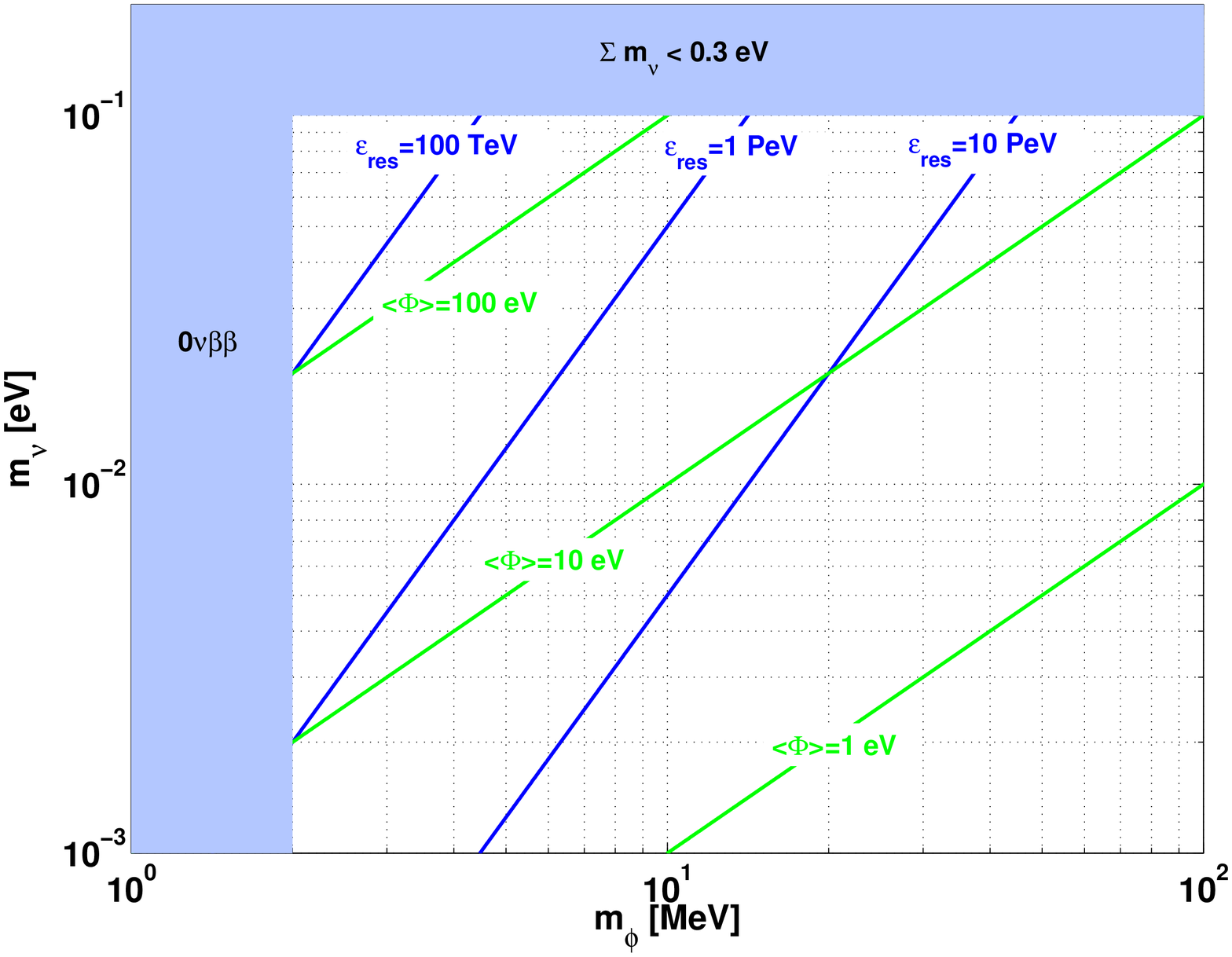}
\end{minipage} 
\caption{{\it Left panel}: Neutrino (thick curves) and cascaded gamma-ray (thin curves) limits on the lifetime of decaying dark matter (90\% CL).  We assume the nondetection in three year observations by IceCube and use measurements of the diffuse gamma-ray background by {\it Fermi}.  Taken from Ref.~\cite{Murase:2012xs}.
{\it Right panel}: Parameter space for secret neutrino-neutrino scattering, where $m_\nu$ is the neutrino mass and $m$ is the mediator mass.  The $\varepsilon_{\rm res}=$const. line represents the resonance condition.  Taken from Ref.~\cite{Blum:2014ewa}.}
\end{figure}

While revealing the origin of PeV neutrinos is an important question, it is interesting to use astrophysical neutrinos as a probe of new physics.  
Although heavy dark matter has been invoked as possible explanations of the IceCube signal, more conservatively, one can put constraints on the lifetime of decaying dark matter or cross section of annihilating dark matter.  The neutrino constraints are powerful at very high energies, while cascade gamma rays give us conservative constraints, as shown in Ref.~\cite{Murase:2012xs} and recently supported by Ref.~\cite{Rott:2014kfa}. 
As another example, we note that PeV neutrino observations allow us to probe nonstandard interactions that cannot be easily probed by laboratory experiments. In view of particle physics, the neutrino sector holds a mystery and neutrino mass mechanisms are related to physics beyond the standard model.  In Ref.~\cite{Blum:2014ewa} (see also Ref.~\cite{Ibe:2014pja}), a low-scale Majorna neutrino mass generation mechanism is considered, where a light scalar and explicit lepton number violation due to its VEV are introduced.  The resulting neutrino self-interactions via a light mediator allow high-energy extragalactic neutrinos to interact with the cosmic neutrino background.  Then, one may expect not only neutrino attenuation but also neutrino cascades, which could be seen in the IceCube data~\cite{Ioka:2014kca,Ng:2014pca}.  In Ref.~\cite{Araki:2014ona}, alternately, the leptonic gauge interaction $L_\mu-L_\tau$ is introduced to explain a possible gap in the IceCube data and the muon anomalous magnetic moment $g_\mu-2$.  Also, the flavor ratio is affected although much more statistics are needed~\cite{Blum:2014ewa}.  Note that flavor studies will allow us to explore various nonstandard properties of neutrinos, including neutrino decay and exotic neutrino mixing via sterile neutrinos propagating in extra dimensions~\cite{Baerwald:2012kc,Pakvasa:2012db,Barger:2014iua}.  
The observation of a possible spectral break or cutoff has also been used to test the Lorentz invariance violation in the neutrino sector~\cite{Diaz:2013wia,Anchordoqui:2014hua,Stecker:2014xja,Learned:2014vya}. 

To address all the key questions, we need more statistics anyway.  And, one of the next goals is to identify individual neutrino sources, even if such sources may not be related to the origin of the diffuse neutrino flux.  This may not be so easy, and we will need next-generation neutrino detectors.  Possibly, rare sources such as bright AGN might be detected especially in the EeV range or stacking analyses would be useful for nearby sources like galaxy clusters.  Searching for Galactic sources should be continued, and KM3Net in the Mediterranean Sea will be useful for this purpose~\cite{Spurio:2014una}.  Transient sources such as GRBs and AGN flares are still promising since atmospheric neutrino and muon backgrounds can be reduced by taking time- and space-coincidence, and we here note that GeV-TeV neutrino searches should be done as well.  We emphasize that multimessenger approaches are crucial in all the cases.  For example, gamma-ray detectors such as HAWC should play a complementary role in searches for Galactic pevatrons.  For transients, it is crucial to have good sky monitors in electromagnetic observations.  Not only gamma-ray monitors but also X-ray monitors and wide optical surveys such as ASAS-SN will be useful for finding hidden treasures.  


\begin{theacknowledgments}
K.M. thanks Markus Ahlers, Kfir Blum, Peter M\'esz\'aros, Eli Waxman, and Walter Winter for useful discussions.  
This work is supported by NASA through Hubble Fellowship Grant No. 51310.01 awarded by the STScI, which is operated by the Association of Universities for Research in Astronomy, Inc., for NASA, under Contract No. NAS 5-26555.  K.M. also acknowledges his host institution, Institute for Advanced Study for continuous support. 
\end{theacknowledgments}



\bibliographystyle{aipproc}   

\bibliography{kmurase.bib}

\IfFileExists{\jobname.bbl}{}
 {\typeout{}
  \typeout{******************************************}
  \typeout{** Please run "bibtex \jobname" to optain}
  \typeout{** the bibliography and then re-run LaTeX}
  \typeout{** twice to fix the references!}
  \typeout{******************************************}
  \typeout{}
 }

\end{document}